\documentclass[letters,useAMS,usenatbib]{mnras}
\usepackage{journals}
\usepackage{graphicx}
\usepackage{color}
\usepackage{times}
\usepackage{hyperref}
\usepackage{amsmath}
\usepackage{bm}
\usepackage{multirow}

\title[SNe Ia in the SFDs of spiral hosts]{Type Ia supernovae in the star formation deserts of spiral host galaxies}
\author[A.~A.~Hakobyan~et~al.]{A.~A.~Hakobyan,$^{1}$\thanks{\fontsize{7.56}{9.1}\selectfont{E-mail:
\href{mailto:artur.hakobyan@yerphi.am}{artur.hakobyan@yerphi.am} (AAH);
\href{mailto:a.karapetyan@yerphi.am}{a.karapetyan@yerphi.am} (AGK)}}
A.~G.~Karapetyan,$^{1}$\textcolor[rgb]{0,0,1}{\footnotemark[1]}
L.~V.~Barkhudaryan,$^{1}$
M.~H.~Gevorgyan$^{1}$
and V.~Adibekyan$^{2,3}$
\\
$^{1}$Center for Cosmology and Astrophysics, Alikhanian National Science Laboratory, 2 Alikhanian Brothers Str., 0036 Yerevan, Armenia\\
$^{2}$Instituto de Astrof\'isica e Ci\^encias do Espa\c{c}o, Universidade do Porto, CAUP, Rua das Estrelas, 4150-762 Porto, Portugal\\
$^{3}$Departamento de F\'isica e Astronomia, Faculdade de Ci\^encias, Universidade do Porto, Rua do Campo Alegre, 4169-007 Porto, Portugal}
\begin{document}

\date{Accepted 2021 May 5. Received 2021 April 10; in original form 2021 January 7}

\pagerange{\pageref{firstpage}--\pageref{lastpage}} \pubyear{2021}

\maketitle

\label{firstpage}

\begin{abstract}
  Using a sample of nearby spiral galaxies hosting 185 supernovae (SNe) Ia,
  we perform a comparative analysis of the locations and light curve decline rates
  $(\Delta m_{15})$ of normal and peculiar SNe~Ia in the star formation deserts (SFDs) and beyond.
  To accomplish this, we present a simple visual classification approach based on
  the UV/H$\alpha$ images of the discs of host galaxies.
  We demonstrate that, from the perspective of the dynamical timescale of the SFD,
  where the star formation (SF) is suppressed by the bar evolution,
  the $\Delta m_{15}$ of SN~Ia and progenitor age can be related.
  The SFD phenomenon gives an excellent possibility to
  separate a subpopulation of SN~Ia progenitors with the ages older than a few Gyr.
  We show, for the first time, that
  the SFDs contain mostly faster declining SNe~Ia $(\Delta m_{15} > 1.25)$.
  For the galaxies without SFDs, the region within the bar radius,
  and outer disc contain mostly slower declining SNe~Ia.
  To better constrain the delay times of SNe~Ia,
  we encourage new studies (e.g. integral field observations)
  using the SFD phenomenon on larger and more robust datasets of
  SNe~Ia and their host galaxies.
\end{abstract}

\begin{keywords}
supernovae: individual: Type Ia -- galaxies: bar -- galaxies: disc --
galaxies: star formation -- galaxies: stellar content.
\end{keywords}

\section{Introduction}
\label{intro}
\defcitealias{2020MNRAS.499.1424H}{H20}

It is believed that the progenitor of Type~Ia supernova (SN~Ia) is
a carbon--oxygen white dwarf (WD) in close binaries,
whose properties and explosion channels are still under debate
\citep[e.g.][]{2018PhR...736....1L}.
SNe~Ia show an important relation between the luminosity at $B$-band maximum
and their light curve (LC) decline rate $\Delta m_{15}$:
faster declining SNe are fainter \citep{1993ApJ...413L.105P}.
The $\Delta m_{15}$ is the difference in magnitudes between the maximum and 15~d after,
and is considered as a practically extinction-independent parameter
\citep[e.g.][hereafter H20]{2020MNRAS.499.1424H}.
Much work has been done to determine the nature of SN~Ia
progenitors by studying the relations between the properties of SNe~Ia
and characteristics of galaxies in which they are discovered
\citep[e.g.][]{2005ApJ...634..210G,2011ApJ...740...92G,2013A&A...560A..66R,
2017ApJ...848...56U,2020ApJ...889....8K}.
In particular, the SN~Ia LC decline rate can be linked to the global age of host galaxy
\citep[e.g.][]{2017ApJ...851L..50S},
which is usually considered as a rough proxy for the SN~Ia delay time
(i.e. time interval between the progenitor formation and its subsequent explosion).
Recently, in \citetalias{2020MNRAS.499.1424H}, we showed that
the correlation between the $\Delta m_{15}$ of normal SNe~Ia and hosts' global age
appears to be due to the superposition of at least two distinct populations of
faster and slower declining SNe~Ia from older and younger stellar populations,
respectively.
For the most common peculiar SNe~Ia, we showed that
91bg-like (subluminous and fast declining) events probably come only from
the old population, while 91T-like (overluminous and slow declining) SNe originate
only from the young population of galaxies.
Such results have been obtained also from more accurate
age estimations of SNe~Ia host populations,
using the local properties for SN sites
\citep[e.g.][]{2013A&A...560A..66R,2019PASA...36...31P,2019ApJ...874...32R}.
Eventually, the SN LC properties, delay time distribution (DTD), and relations with
other host characteristics allow to constrain the SN~Ia progenitor scenarios
(see discussion in \citetalias{2020MNRAS.499.1424H}).

In this \emph{Letter}, for the first time, we link the $\Delta m_{15}$ of
SN~Ia with the progenitor age from the perspective of
star formation desert (SFD) phenomenon.
In short, the SFD, observed in some spiral galaxies
\citep[e.g.][]{2015MNRAS.450.3503J,2018MNRAS.474.3101J},
is a region swept up by a strong bar with almost no recent star formation (SF)
on both sides of the bar.
There are increasing evidences in observations and simulations that
SFD consists of old stars, and the quenching of SF in this region
was due to the bar formation \citep[e.g.][]{2019MNRAS.489.4992D,2020A&A...644A..79G},
which dynamically removed gas from SFD
over a timescale of $\sim2$~Gyr \citep[e.g.][]{2019MNRAS.489.4992D}.
The bar can show SF through its length, or
SF can be found only at the bar ends,
or the entire bar might not show SF \citep[e.g.][]{2020A&A...644A..38D}.
Some bars might even dissolve during the evolution \citep[e.g.][]{2004ApJ...604..614S},
leaving the central SFD in galactic disc.
On the other hand, it can be considered that the SFD is practically not contaminated by
the radial migration of young stars from the outer disc
\citep[e.g.][]{2018MNRAS.481.1645M}.
Therefore, from the dynamical age-constrain of SFD ($\gtrsim2$~Gyr),
we consider that the DTD of its SNe~Ia is truncated on the younger side,
starting from a few Gyr, in comparison with those outside the SFD,
where mostly young/prompt SNe~Ia occur \citep[delay time of $\sim 500$~Myr,][]{2009ApJ...707...74R}.
Given this, and if the progenitor's age is the main driver of the decline rate,
the SNe~Ia discovered in the SFDs should have faster declining LCs.
In this study, we simply demonstrate the validity of
this assumption according to the picture briefly described above,
which provides an excellent new opportunity to constrain the nature of SN~Ia progenitors.

\section{Sample selection and reduction}
\label{samplered}

We selected the sample for our study
from a well-defined sample of \citetalias{2020MNRAS.499.1424H}, which includes data
on the spectroscopic subclasses of
nearby ($\leq$150~Mpc) SNe~Ia (normal, 91T-, 91bg-like, etc.)
and their $B$-band LC decline rates $(\Delta m_{15})$,
as well as homogeneous data on the host galaxies (distance, corrected
$ugriz$ magnitudes, morphological type, bar detection, etc).
The SFDs are observed in some barred Sa--Scd galaxies
\citep[e.g.][]{2015MNRAS.450.3503J,2018MNRAS.474.3101J},
therefore we restricted the morphologies of SNe hosts to
the mentioned types, with barred and unbarred counterparts.
We also ignored hosts with strong morphological disturbances,
which may add undesirable projection effects and complicate
the assignment of an SN~Ia to the SFD.

As seen in \citet{1997ApJ...483L..29W,2015MNRAS.448..732A,2016MNRAS.456.2848H},
the vast majority of SNe~Ia in spiral galaxies belong to the disc,
rather than the bulge (spherical) component.
Here, we checked this observational fact for the most common SN~Ia subclasses separately.
If SNe~Ia belong mostly to the disc component, where SFD can be located,
one would expect that the distributions of projected and $R_{25}$-normalized galactocentric
distances\footnote{{\footnotesize Normalized to the $g$-band $25^{\rm th}$ magnitude isophotal
semimajor axis of host galaxy $(R_{25}=D_{25}/2)$.}}
of SNe~Ia along major $(|U|/R_{25})$ and minor $(|V|/R_{25})$ axes
would be different, being distributed closer to the major axis
\citep[i.e. smaller $\langle |V|/R_{25} \rangle$ in comparison with $\langle |U|/R_{25} \rangle$, see][for more details]{2016MNRAS.456.2848H}.
Table~\ref{UVR25disKSAD} shows the results of the two-sample Kolmogorov--Smirnov (KS) and
Anderson--Darling (AD) tests on the comparison of the major versus minor axes
distributions of 238 SNe~Ia (based on a subsample from \citetalias{2020MNRAS.499.1424H}).
The $P$-values of the tests suggest that the SNe~Ia distribution along the major axis is inconsistent
with that along the minor axis in Sa--Scd host galaxies with different inclinations,
showing that the SN~Ia subclasses in these hosts originate mostly from the disc population.

It should be noted that, because of the absorption and projection effects in the discs,
the SFDs are observed in some spiral galaxies only with low/moderate inclinations
\citep[e.g.][]{2015MNRAS.450.3503J,2018MNRAS.474.3101J}.
Therefore, we also limited our host galaxy sample to inclinations $i < 70^\circ$.
In total, there are 185 normal, 91T- and 91bg-like SNe~Ia
meeting the above criteria, of which 79 and 106 events
have barred and unbarred hosts, respectively.
These SNe~Ia are discovered in 180 host galaxies,
five of which host two events in each.

\begin{figure*}
\begin{center}$
\begin{array}{@{\hspace{0mm}}c@{\hspace{0mm}}c@{\hspace{0mm}}c@{\hspace{0mm}}c@{\hspace{0mm}}c@{\hspace{0mm}}}
\includegraphics[width=0.18\hsize]{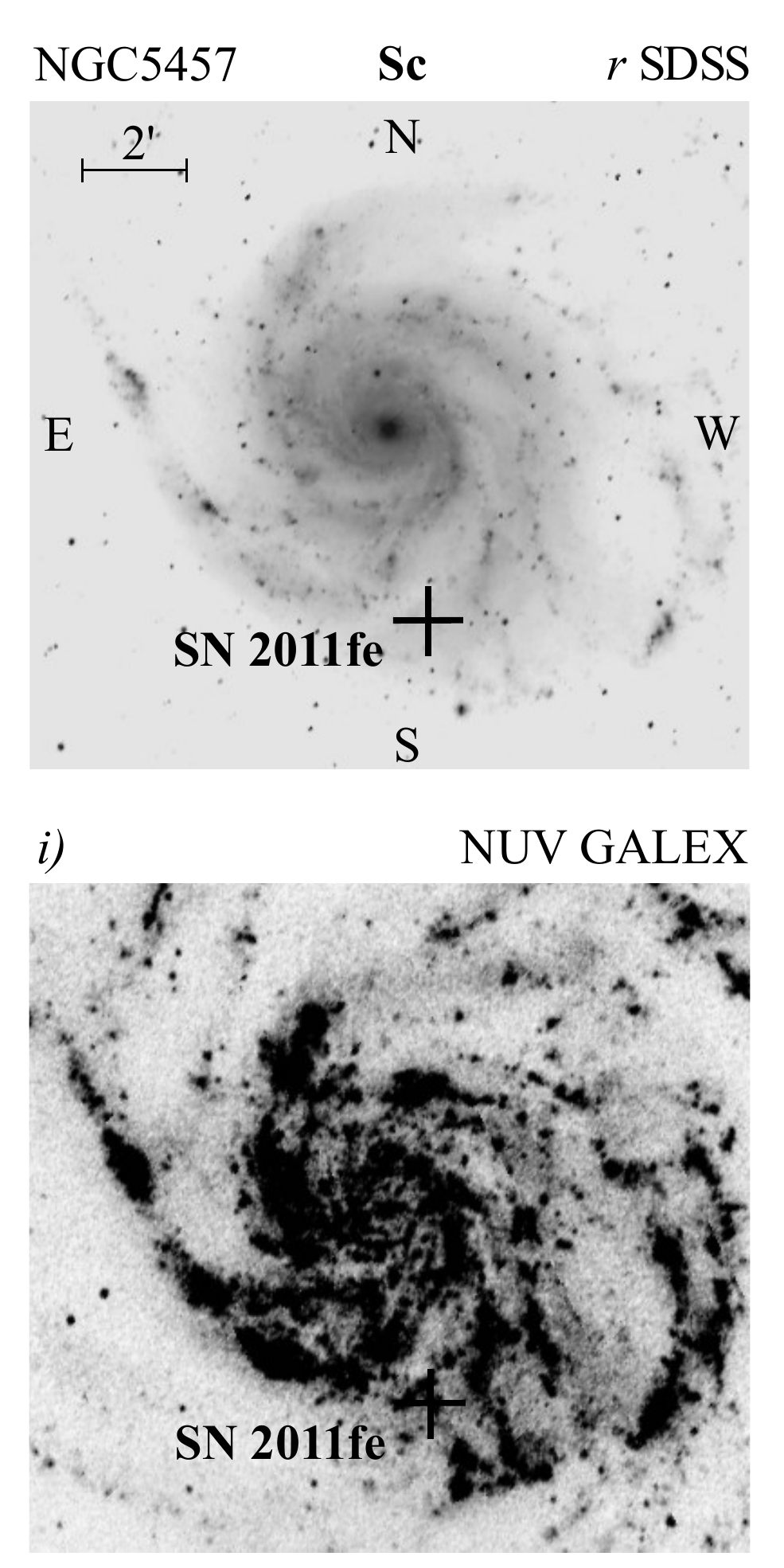} &
\includegraphics[width=0.18\hsize]{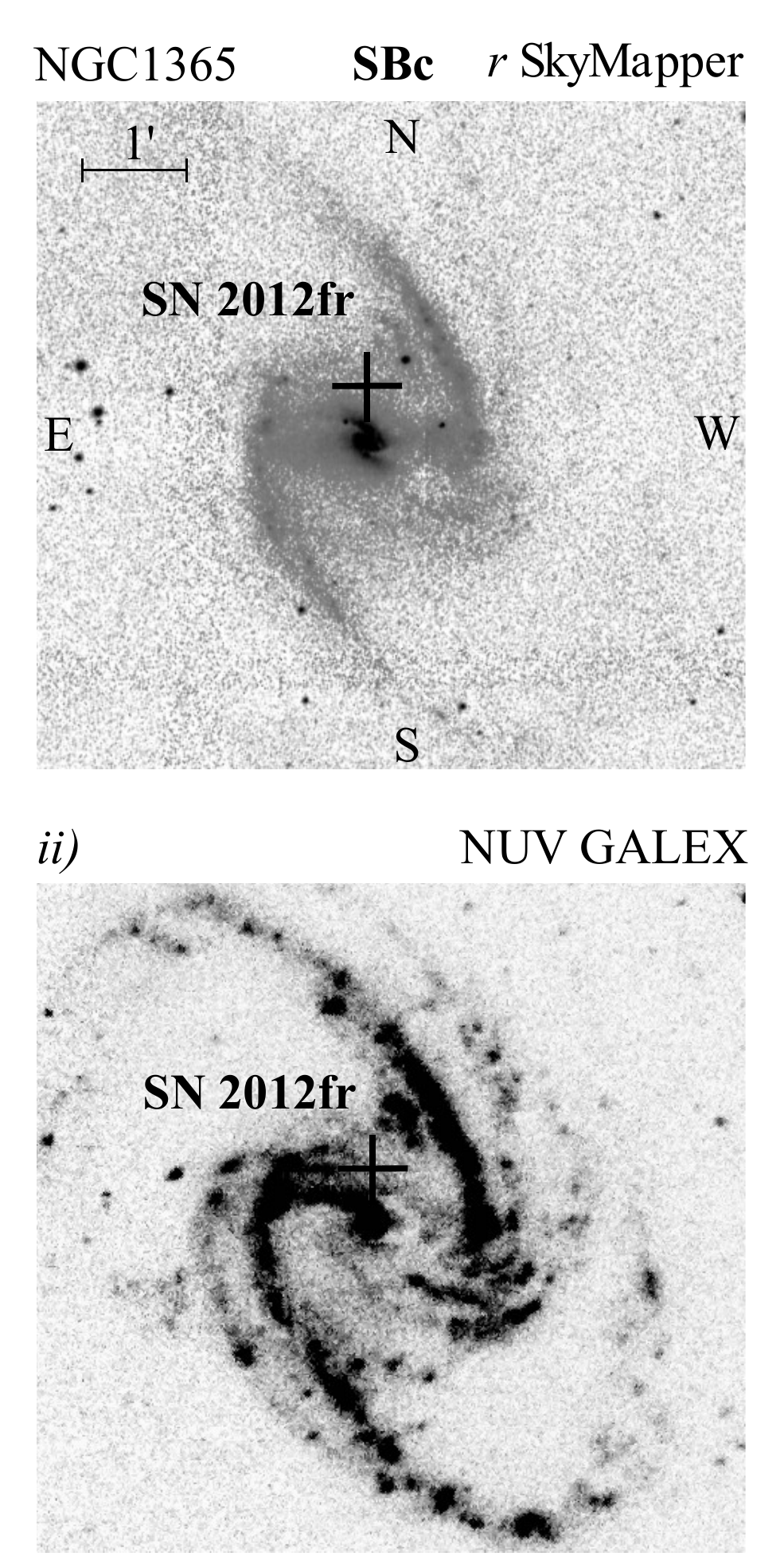} &
\includegraphics[width=0.18\hsize]{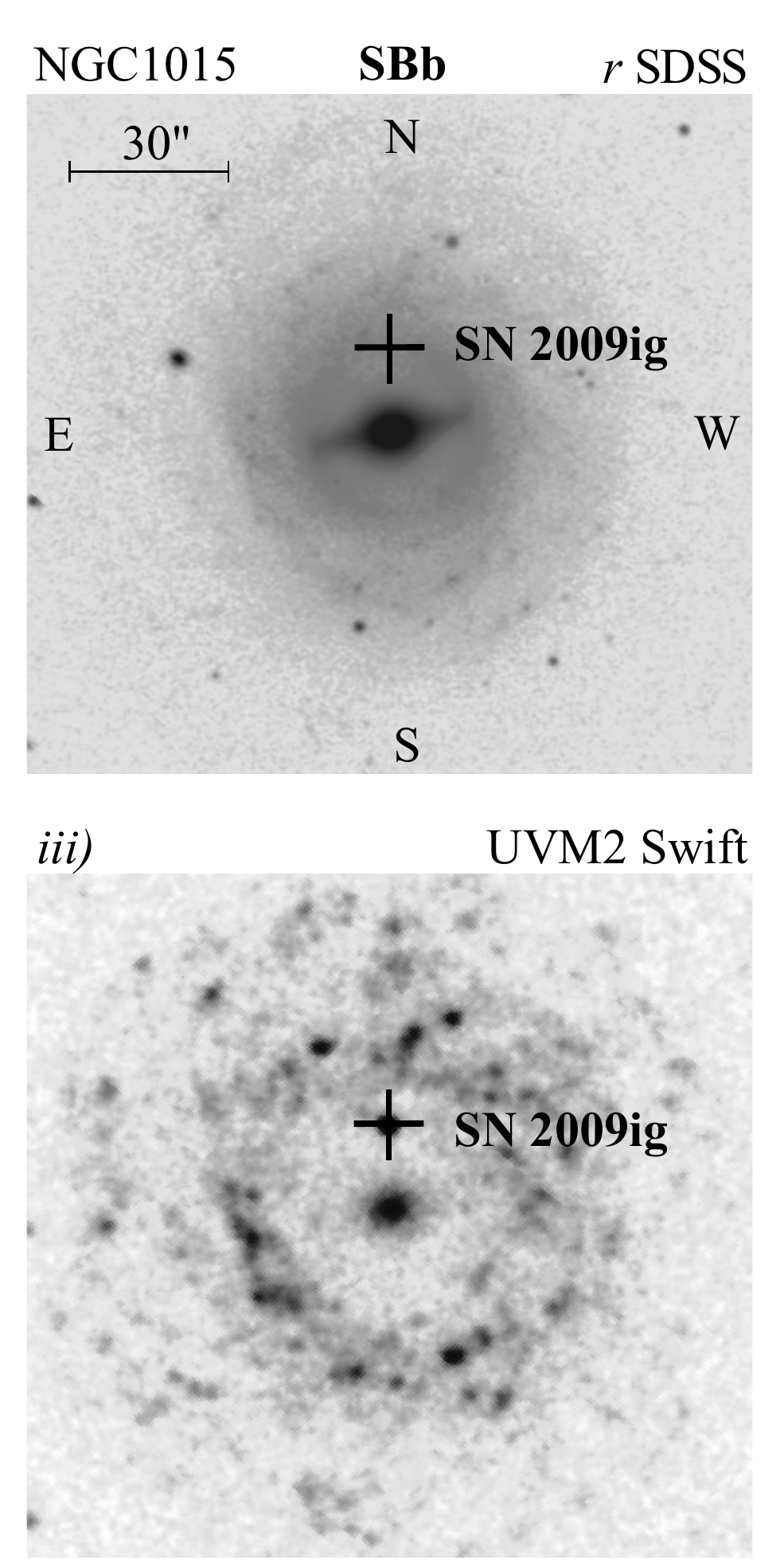} &
\includegraphics[width=0.18\hsize]{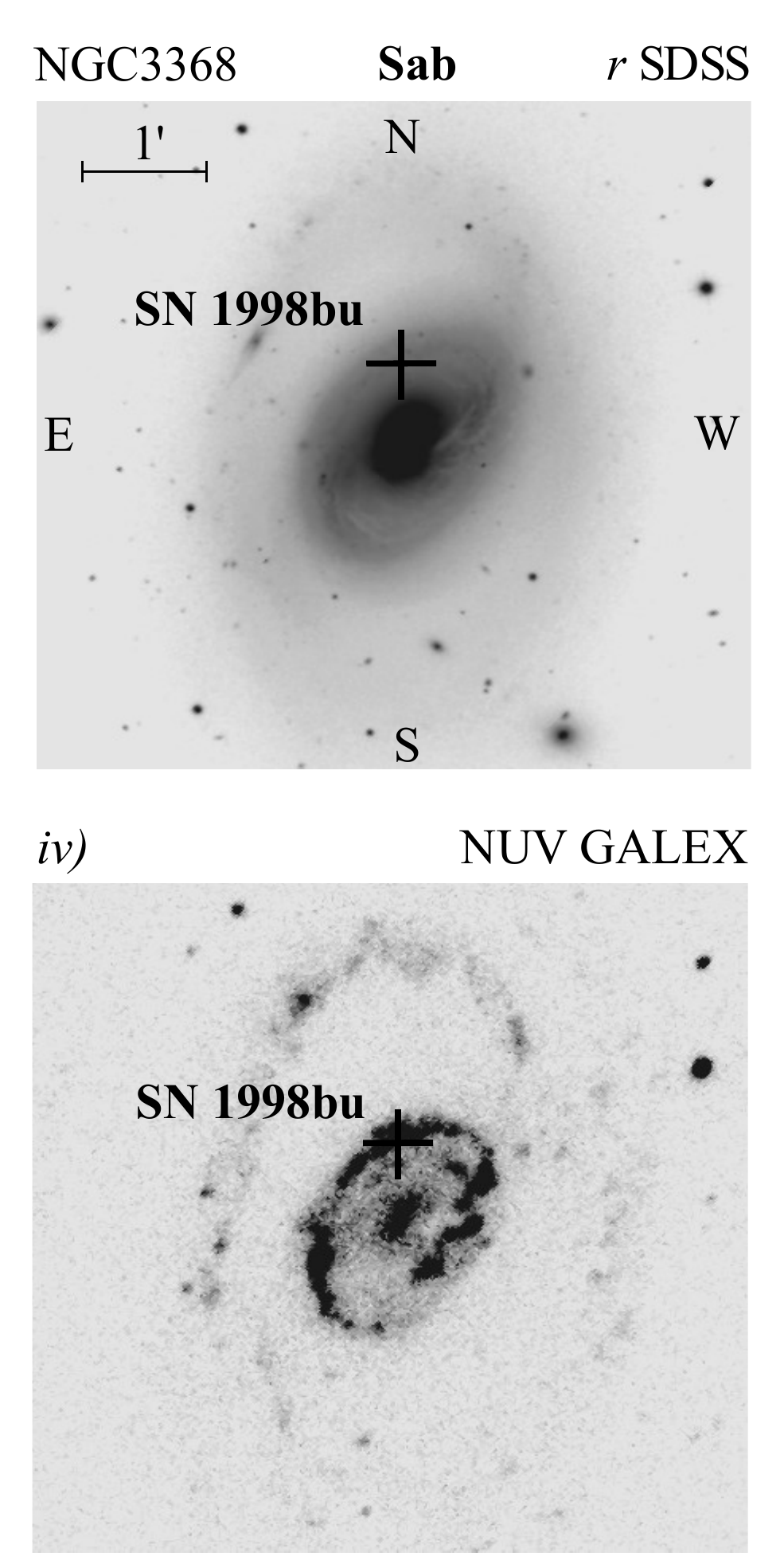}
\end{array}$
\end{center}
\caption{Optical (top) and UV images (bottom) representing examples of SN~Ia hosts
         with different SF classes of their discs.
         Galaxies' identifiers, morphologies, and discs' classes are listed at the top.
         Classes \emph{i} and \emph{ii} do not have SFD,
         while classes \emph{iii} and \emph{iv} have SFD.
         SN~Ia names/positions are also shown.}
\label{SFSFDexamples}
\end{figure*}

For these host galaxies,
we used archival Galaxy Evolution Explorer (GALEX) far- and near-UV
\citep[][]{2005ApJ...619L...1M},
\emph{Swift} UV \citep[][]{2005SSRv..120...95R},
and available H$\alpha$ images \citep[e.g.][]{2018A&A...609A.119S}
to visually classify the morphology
of their ionized discs into four SF classes:
\emph{i)} SF is distributed along the entire length of unbarred disc, from the center to the edge
(97 SNe~Ia hosts);
\emph{ii)} like in the first case, but for barred disc, SF along the bar, without SFD
(36 objects);
\emph{iii)} SF along the bar, or SF might occur only at the bar ends, with SFD
(43 objects);
\emph{iv)} SF is distributed along the unbarred disc, except the central SFD
(9 objects).
In all the cases, the circumnuclear SF is also possible.
The $r$-band and UV images representing
the classes\footnote{{\footnotesize With some differences,
a similar visual classification is also proposed by \citet{2020A&A...644A..38D}.}}
of galaxies can be found in Fig.~\ref{SFSFDexamples}.
Note that the cosmic surface brightness dimming is insignificant
for our galaxy sample, since hosts' $z\lesssim 0.036$
$(\langle z\rangle=0.017\pm0.009)$.

\begin{table}
  \centering
  \begin{minipage}{84mm}
  \caption{Comparison of the projected and normalized distributions of the
           SN~Ia subclasses along major $(U)$ and minor $(V)$ axes of
           Sa--Scd hosts.}
  \tabcolsep 4.3pt
  \label{UVR25disKSAD}
    \begin{tabular}{lrcccrr}
    \hline
  \multicolumn{1}{l}{SN} & \multicolumn{1}{c}{$N_{\rm SN}$} &\multicolumn{1}{c}{Subsample~1} &
  \multicolumn{1}{c}{vs} & \multicolumn{1}{c}{Subsample~2} & \multicolumn{1}{c}{$P_{\rm KS}^{\rm MC}$} &
  \multicolumn{1}{c}{$P_{\rm AD}^{\rm MC}$}\\
  && \multicolumn{1}{c}{$\langle |U|/R_{25} \rangle$} && \multicolumn{1}{c}{$\langle |V|/R_{25} \rangle$} && \\
  \hline
    Normal & 196 & 0.32$\pm$0.02 & vs & 0.20$\pm$0.02 & $<$\textbf{0.001} & $<$\textbf{0.001}\\
    91T & 27 & 0.38$\pm$0.06 & vs & 0.20$\pm$0.03 & \textbf{0.018} & \textbf{0.021}\\
    91bg & 15 & 0.34$\pm$0.06 & vs & 0.18$\pm$0.06 & \textbf{0.020} & \textbf{0.037}\\
  \hline
  \end{tabular}
  \small
  \parbox{\hsize}{\emph{Notes:} The $P_{\rm KS}^{\rm MC}$ $(P_{\rm AD}^{\rm MC})$
                  is the two-sample KS (AD) test
                  probability that the distributions are drawn from the same parent sample,
                  using a Monte Carlo (MC) simulation with $10^5$ iterations
                  as explained in \citetalias{2020MNRAS.499.1424H}.
                  The respective mean values and standard errors are listed.
                  The statistically significant differences $(P \leq 0.05)$ between the distributions
                  are highlighted in bold.}
  \end{minipage}
\end{table}

Based on the optical $g$-band images, we measured bar radii
of host galaxies using ellipse fitting to the bar isophotes with maximum ellipticity
(see \citealt{2016A&A...587A.160D} and references therein,
for more details on the bar radius measurement method).
Then we deprojected each bar radius for host inclination
and normalized it to the disc radius, i.e.
$\widetilde{r}_{\rm bar} = R_{\rm bar}/R_{25}$.
For unbarred hosts with the central SFD
(class~\emph{iv} in Fig.~\ref{SFSFDexamples}), we used the UV images
to roughly estimate the radii of SFDs $(\widetilde{r}_{\rm SFD} = R_{\rm SFD}/R_{25})$,
where almost no UV fluxes are detected.
Note that for our sample $\langle\widetilde{r}_{\rm SFD}\rangle \approx
\langle\widetilde{r}_{\rm bar}\rangle = 0.30$.
For further simplicity, we define a demarcation radius as:\\
$\widetilde{r}_{\rm dem} = \begin{cases}
    \widetilde{r}_{\rm bar}, & {\rm for} \, ii \, {\rm and} \, iii {\rm \, disc \, classes}, \\
    \widetilde{r}_{\rm SFD}, & {\rm for} \, iv {\rm \, class}.
\end{cases}$\\
For SNe~Ia, we deprojected and normalized their galactocentric distances as well,
i.e. $\widetilde{r}_{\rm SN} = R_{\rm SN}/R_{25}$
\citep[see][]{2016MNRAS.456.2848H}.

Based on the host disc classification and the definition of demarcation radius,
we grouped SNe according to their locations as follows:
97 SNe~Ia are found in the disc of galaxies without a bar or SFD;
61 SNe are in the outer disc of hosts, which have either a bar or SFD;
13 SNe are found in bar or star-forming regions inside $\widetilde{r}_{\rm dem}$;
and 14 SNe~Ia are in SFD.
Table~\ref{SFSFDRSNRbar} displays the distribution of the SN~Ia subclasses
according to their locations in the SFD\footnote{{\footnotesize Although
detailed measurements of the underlying fluxes at SNe locations
are beyond the scope of this \emph{Letter}, nevertheless, for our confidence,
we checked the vicinity of 14 SNe~Ia in the SFDs simply using different apertures
on the fits images, and found no detectable underlying UV/H$\alpha$ fluxes.}}
or beyond.

\begin{table}
  \centering
  \begin{minipage}{84mm}
  \caption{Numbers of the SN~Ia subclasses
           according to their locations in the SFD of
           Sa--Scd hosts or beyond.}
  \tabcolsep 6.9pt
  \label{SFSFDRSNRbar}
    \begin{tabular}{lccccc}
    \hline
  \multicolumn{1}{l}{SNe~Ia~in} & \multicolumn{1}{c}{class~\emph{i}~disc} & \multicolumn{1}{c}{outer~disc} & \multicolumn{1}{c}{bar/SF} & \multicolumn{1}{c}{SFD} & \multicolumn{1}{c}{All} \\
  $\widetilde{r}_{\rm SN}$ & $>$0 & $\geq$$\widetilde{r}_{\rm dem}$ & $<$$\widetilde{r}_{\rm dem}$ & $<$$\widetilde{r}_{\rm dem}$ & \\
  \hline
    Normal & 75 & 52 & 12 & 12 & 151 \\
    91T & 19 & 4 & 1 & 0 & 24 \\
    91bg & 3 & 5 & 0 & 2 & 10 \\
    All & 97 & 61 & 13 & 14 & 185 \\
  \hline
  \end{tabular}
  \end{minipage}
\end{table}

\section{Results and discussion}
\label{RESults}

With the aim of linking the $\Delta m_{15}$ of SN~Ia with the progenitor age,
we study the SN decline rates that
exploded in SFDs and other regions of hosts.
In addition, we compare the SN galactocentric distances
between the spectroscopic subclasses,
and check the possible correlations between the $\Delta m_{15}$ and galactocentric distances.

\subsection{SNe~Ia in the SFDs and beyond}
\label{RESults1}

\begin{figure}
\begin{center}$
\begin{array}{@{\hspace{0mm}}c@{\hspace{0mm}}}
\includegraphics[width=0.90\hsize]{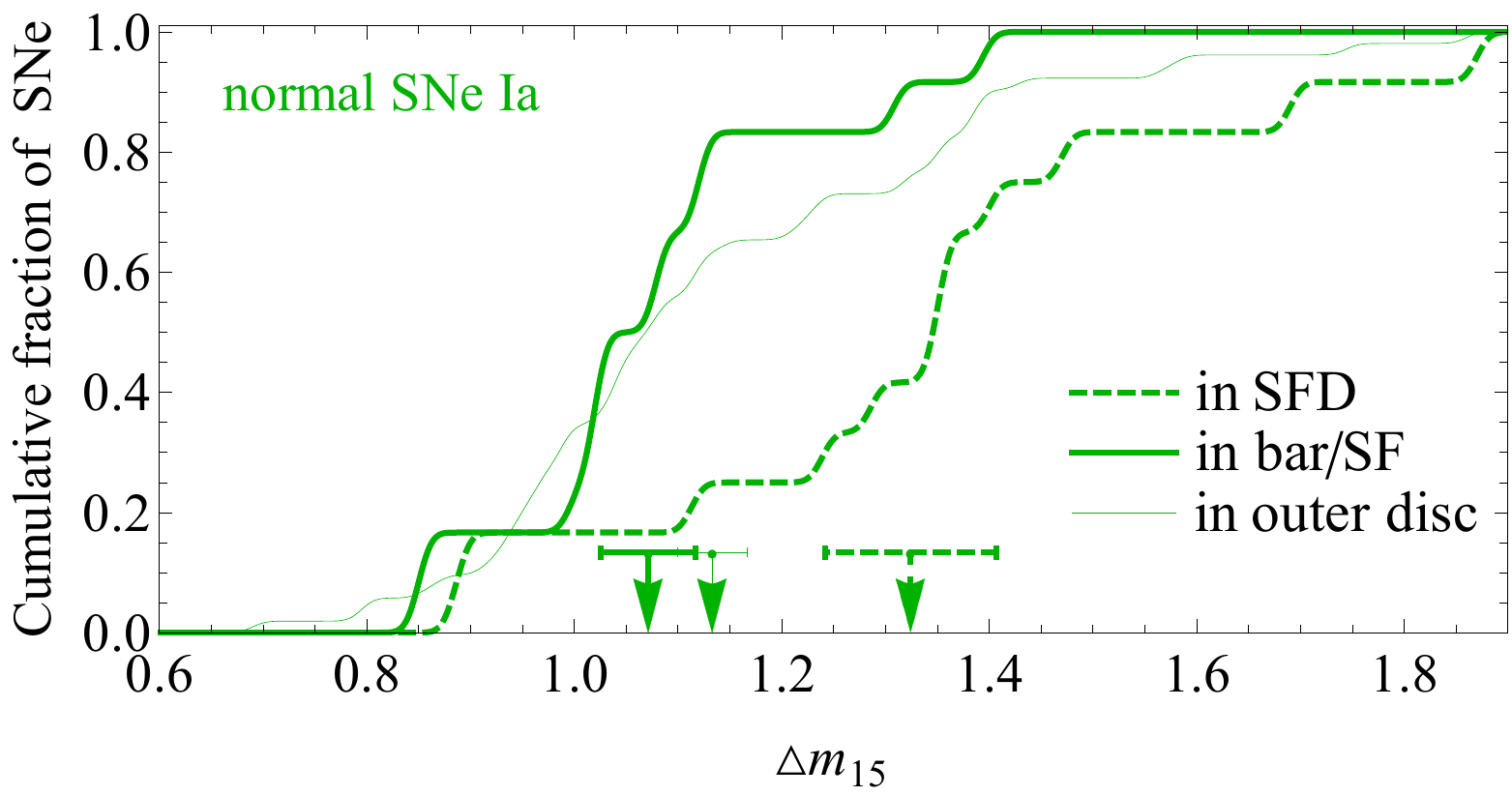}\\
\includegraphics[width=0.93\hsize]{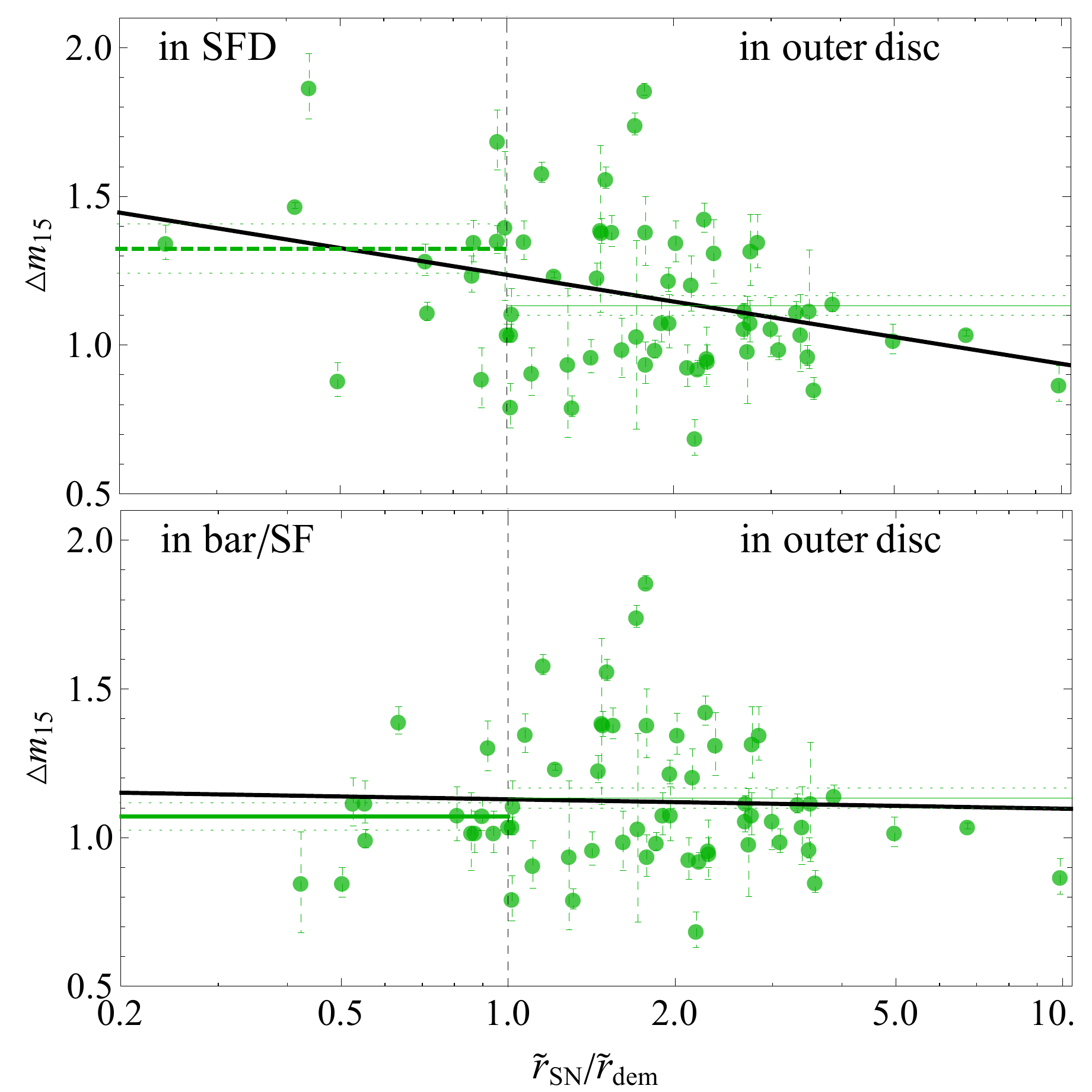}
\end{array}$
\end{center}
\caption{Upper panel: cumulative $\Delta m_{15}$ distributions for normal SNe~Ia
         inside (in SFD and bar/SF) and outside the demarcation radius
         (in outer disc).
         Bottom panel: variation of the $\Delta m_{15}$
         as a function of $\widetilde{r}_{\rm dem}$-normalized galactocentric distance,
         split between different SN locations.
         The best-fits are shown by thick (black) solid lines.
         The vertical dashed line indicates the location of radial demarcation (see the text for details).
         The mean values (with standard errors) of the distributions are shown
         by arrows (with error bars) in the upper panel, and by horizontal lines in the bottom panel.}
\label{dm15rsnrdemcum}
\end{figure}

To link the LC properties of SN~Ia with the progenitor age
from the perspective of the dynamical age-constrain of SFD,
in Table~\ref{dm15inoutrdem},
we compare the $\Delta m_{15}$ distribution of normal SNe~Ia in the SFD
with that in the bar/SF (see also the upper panel of Fig.~\ref{dm15rsnrdemcum}).
The KS and AD tests show that these distributions are significantly different.
Normal SNe~Ia that are in the SFD, dominated by the old population
\citep[$\gtrsim$ 2~Gyr;][]{2019MNRAS.489.4992D},
have, on average, faster declining LCs compared to those located in the bar/SF,
where UV/H$\alpha$ fluxes are observed \citep[i.e. age $\lesssim$ a few 100~Myr;][]{1998ARA&A..36..189K}.

\begin{table}
  \centering
  \begin{minipage}{84mm}
  \caption{Comparison of the $B$-band $\Delta m_{15}$ distributions between normal SNe~Ia
           in different locations (as described in Table~\ref{SFSFDRSNRbar}).}
  \tabcolsep 2.1pt
  \scriptsize
  \label{dm15inoutrdem}
    \begin{tabular}{cccccccrr}
    \hline
  \multicolumn{3}{c}{------------ Subsample~1 ------------} & \multicolumn{1}{c}{vs} & \multicolumn{3}{c}{--------- Subsample~2 ---------} & \multicolumn{1}{c}{$P_{\rm KS}^{\rm MC}$} & \multicolumn{1}{c}{$P_{\rm AD}^{\rm MC}$}\\
  \multicolumn{1}{c}{SN~in} & \multicolumn{1}{c}{$N_{\rm SN}$} & \multicolumn{1}{c}{$\langle \Delta m_{15} \rangle$} && \multicolumn{1}{c}{SN~in} & \multicolumn{1}{c}{$N_{\rm SN}$} & \multicolumn{1}{c}{$\langle \Delta m_{15} \rangle$} && \\
  \hline
     SFD & 12 & 1.32$\pm$0.08 & vs & bar/SF& 12 & 1.07$\pm$0.05 & \textbf{0.005} & \textbf{0.020}\\
     SFD & 12 & 1.32$\pm$0.08 & vs & outer~disc & 52 & 1.13$\pm$0.03 & \textbf{0.009} & \textbf{0.029}\\
     bar/SF & 12 & 1.07$\pm$0.05 & vs & outer~disc & 52 & 1.13$\pm$0.03 & 0.660 & 0.682\\
     SFD$+$bar/SF & 24 & 1.20$\pm$0.05 & vs & outer~disc & 52 & 1.13$\pm$0.03 & 0.445 & 0.477\\
     inner~class~\emph{i}~disc & 17 & 1.19$\pm$0.05 & vs & outer~class~\emph{i}~disc & 58 & 1.11$\pm$0.02 & 0.517 & 0.253\\
  \hline
  \end{tabular}
  \small
  \parbox{\hsize}{\emph{Notes:} Since, the
                  $\langle\widetilde{r}_{\rm dem}\rangle = 0.30$
                  for class \emph{ii-iv} discs, we define inner and outer
                  class \emph{i} discs when $\widetilde{r}_{\rm SN}<0.30$ and
                  $\geq0.30$, respectively.
                  The explanations for $P$-values are the same as in Table~\ref{UVR25disKSAD}.}
  \end{minipage}
\end{table}

Table~\ref{dm15inoutrdem} also shows that
the $\Delta m_{15}$ distribution of normal SNe~Ia that are in the outer disc population
is consistent with that in the bar/SF and
inconsistent with that in the SFD (see also Fig.~\ref{dm15rsnrdemcum}).
Interestingly, any inconsistency vanishes when we combine the bar/SF and SFD subsamples
and compare the LC decline rates with those in the
outer disc population (Table~\ref{dm15inoutrdem}).
This suggests that the discs of Sa--Scd hosts are indeed outnumbered by normal SNe~Ia
with slower declining LCs
(e.g. $\Delta m_{15} < 1.25$, outside the $\widetilde{r}_{\rm dem}$ in Fig.~\ref{dm15rsnrdemcum})
whose progenitor ages peak below 1~Gyr, corresponding to the young/prompt SNe~Ia
\citep[e.g.][]{2014MNRAS.445.1898C}.

In addition, even for discs of class~\emph{i} (without demarcation radius),
the KS and AD tests, in Table~\ref{dm15inoutrdem},
show that the $\Delta m_{15}$ distributions are consistent for
normal SNe~Ia in the inner and outer discs,
excluding a radial dependency of $\Delta m_{15}$ (see also Section~\ref{RESults2}).
For class~\emph{i} discs, the $\langle \Delta m_{15} \rangle$ values are sufficiently consistent with
the same values in the corresponding radial intervals for hosts
having a demarcation radius (see Table~\ref{dm15inoutrdem}).
Thus, the SFD phenomenon gives an excellent possibility to separate
a subpopulation of normal SNe~Ia with old progenitors
from a general population of host galactic disc,
which contains both young and old progenitors.
On average, the LCs of this SN~Ia subpopulation decline faster,
whose DTD is most likely truncated on the younger side,
starting from a several Gyr ($\gtrsim$ 2~Gyr).

These results are qualitatively agree with the theoretical predictions.
In particular, for sub-Chandrasekhar mass ($M_{\rm Ch} \approx 1.4 M_{\odot}$) explosion models
in double WD systems, the luminosity of SN~Ia is directly related to the exploding WD's mass,
which decreases with age
\citep[e.g.][]{2010ApJ...714L..52S,2017MNRAS.470..157B,2017ApJ...851L..50S,2021ApJ...909L..18S}.
This is because WD's mass is directly linked to the main-sequence (MS) mass
of the progenitor star, which is in turn related to the MS lifetime.
Therefore, older stellar populations would host less luminous SNe~Ia,
i.e. faster declining events \citep[e.g.][]{2017ApJ...851L..50S}.
Note that, we prefer sub-$M_{\rm Ch}$ explosion models, because different mechanisms of
the $M_{\rm Ch}$ explosions do not reproduce the observed distribution in
the luminosity--decline rate relation for various SN~Ia subclasses
(e.g. \citealt{2018PhR...736....1L} and references therein).

Despite the small number statistics of peculiar 91T- and 91bg-like SNe,
Table~\ref{SFSFDRSNRbar} shows that the old SFDs of Sa--Scd galaxies
host along with faster declining normal SNe~Ia also two 91bg-like (fast declining) events.
While the bar/SF hosts along with slower declining normal events
also one 91T-like (slow declining) SNe.
Outer disc population hosts all the SN~Ia subclasses (see Table~\ref{SFSFDRSNRbar}).
The latter is also correct for the entire class~\emph{i}~disc.
These results can be explained from the perspective of the SFD's properties
in addition to the previously known relations between
the SNe~Ia and the global (or SN local) properties of their hosts.

In particular, the discovery of 91bg-like events
(progenitor age is greater than several Gyr, e.g.
\citealt{2017NatAs...1E.135C,2019PASA...36...31P,2019MNRAS.490..718B};
\citetalias{2020MNRAS.499.1424H}) and
a population of faster declining normal SNe~Ia
in the SFDs can be explained within the scenario of SF suppression by bar,
where the SFDs of galaxies show
a sharp truncation in SF histories and contain mostly
old stellar population of several Gyr
\citep[$\gtrsim$ 2~Gyr;][]{2019MNRAS.489.4992D}.
On the other hand, the discovery of 91T-like SNe
\citep[progenitor age is less than a Gyr, e.g.][]{2004MNRAS.350.1301H,
2013MNRAS.429.1425R,2015ApJ...805..150F}
and a population of slower declining normal events
in the bar/SF (Tables~\ref{SFSFDRSNRbar}-\ref{dm15inoutrdem}),
can be explained in the context of SF suppression scenario, where the recently formed bar,
within the $\sim1.5$~Gyr timescale, has not yet completely removed the gas
and quenched ongoing SF inside the demarcation radius \citep[][]{2019MNRAS.489.4992D}.
Recall that in the bar/SF regions, the UV fluxes are observed
that trace the SF up to a few 100~Myr \citep[][]{1998ARA&A..36..189K}.

The outer disc of Sa--Scd galaxies (or entire class~\emph{i} disc),
contains stellar populations of all ages
\citep[e.g.][]{2015A&A...581A.103G}.
Therefore, the appearance of all the SN~Ia subclasses
in this region is not unexpected (Table~\ref{SFSFDRSNRbar}).
Note that the results in Table~\ref{dm15inoutrdem} remain statistically unchanged
when we combine \citep[following][]{2017ApJ...851L..50S}
normal, 91T-, and 91bg-like SNe together.

To test different galaxy properties that could affect
the results in Tables~\ref{SFSFDRSNRbar}-\ref{dm15inoutrdem},
we compare the distributions of morphologies, masses, colours, and
ages (available in \citetalias{2020MNRAS.499.1424H})
of classes \emph{ii}-\emph{iv} hosts with and without SFD.
The KS and AD tests show that the global parameters of hosts
are not statistically different $(P > 0.1)$,
thus could not be the main drivers behind our results.
On the other hand, the bar/SF regions have higher surface brightness
and dust content in comparison with the SFDs, and therefore the discovery of
intrinsically faint (faster declining) SNe in the bar/SF can be complicated,
biasing the statistical results in Table~\ref{dm15inoutrdem}.
However, this does not affect the result that the SFD's SNe~Ia are
mostly faster declining (fainter) events.

Let us now briefly address the possible effects of the progenitor metallicity,
which theoretically might cause a variation in the SN~Ia LC properties.
The mean radial metallicity profile of Sa--Scd galaxies declines from solar to
$\sim0.3$~dex below solar from the galactic centre up to the disc end, respectively
\citep[e.g.][]{2015A&A...581A.103G}.
On the other hand, the simulation by \citet{2013A&A...553A.102D}
shows that the metallicity on both sides of the bar,
i.e. in SFD, is only $\sim0.15$~dex below solar.
For any progenitor model, such metallicity variations can account
for less than 0.2~mag in SN~Ia maximum brightness and about 0.1~mag in $\Delta m_{15}$
\citep*[e.g.][]{2003ApJ...590L..83T,2009Natur.460..869K},
which is not enough to be the main reason for the
observed differences in $\Delta m_{15}$ values in SFD and beyond
(Fig.~\ref{dm15rsnrdemcum} and Table~\ref{dm15inoutrdem}).
Thus, our results support earlier suggestions that
the progenitor age is most probably the decisive factor
shaping the observed distribution of SN~Ia decline rates
\citep[e.g.][]{2005ApJ...634..210G}.
Nevertheless, we would like to stress that the discussed effect of metallicity
is heavily based on a very limited number of models.
Therefore, further modelling of the impact of metallicity on the LC properties
of SNe Ia would help to place our findings in context.

\subsection{The radial distribution of SNe~Ia}
\label{RESults2}

In spiral discs, a radial gradient of physical properties of stellar population
\citep[e.g. age gradient;][]{2015A&A...581A.103G} might be a useful tool
and has been used in the past to probe the possible dependencies
of SNe~Ia decline rates on their galactocentric distance
\citep[e.g.][]{2005ApJ...634..210G,2012ApJ...755..125G,2017ApJ...848...56U}.
However, in these studies, the authors were unable to find a significant correlation
between the decline rate and $\widetilde{r}_{\rm SN}$,
which is correct also for our sample (Table~\ref{SpRcordm15vsR}).
Moreover, the radial distributions of peculiar (extreme decliners) and normal SNe~Ia
in Sa--Scd galaxies are consistent with one another $(P > 0.2)$ \citep[e.g.][]{2016AstL...42..495P}.
Note that these results remain statistically insignificant $(P > 0.1)$
when we perform the same tests after separating the hosts into barred/unbarred, and early/late-types.
The $\Delta m_{15}$ is not correlated with $\widetilde{r}_{\rm SN}$
$(P > 0.4)$ also for the class~\emph{i} disc only,
where no bar/SFD phenomena are observed.

\begin{table}
  \centering
  \begin{minipage}{84mm}
  \caption{Spearman's rank correlation test for the $B$-band $\Delta m_{15}$ values
           of the SN~Ia subclasses versus their
           galactocentric distances.}
  \tabcolsep 1.3pt
  \label{SpRcordm15vsR}
  \begin{tabular}{lcrrr}
  \hline
    \multicolumn{1}{l}{SN}&
    \multicolumn{1}{c}{Galactocen. dist.}&\multicolumn{1}{c}{$N_{\rm SN}$}&
    \multicolumn{1}{c}{$r_{\rm s}$}&
    \multicolumn{1}{c}{$P_{\rm s}^{\rm MC}$}\\
  \hline
    Normal & $\widetilde{r}_{\rm SN}$ & 151 & $-$0.070 & 0.394\\
    91T & $\widetilde{r}_{\rm SN}$ & 24 & 0.087 & 0.686\\
    91bg & $\widetilde{r}_{\rm SN}$ & 10 & 0.503 & 0.138\\
    Normal~(SFD$+$outer disc) & $\widetilde{r}_{\rm SN}/\widetilde{r}_{\rm dem}$ & 64 & $-$0.280 & \textbf{0.025}\\
    Normal~(bar/SF$+$outer disc) & $\widetilde{r}_{\rm SN}/\widetilde{r}_{\rm dem}$ & 64 & 0.019 & 0.879\\
    Normal~(SFD$+$bar/SF$+$outer disc) & $\widetilde{r}_{\rm SN}/\widetilde{r}_{\rm dem}$ & 76 & $-$0.147 & 0.206\\
  \hline
  \end{tabular}
  \small
  \parbox{\hsize}{\emph{Notes:} Coefficient $r_{\rm s}$ $(\in[-1;1])$
                  is a measure of rank correlation.
                  The variables are not independent when $P \leq 0.05$.
                  The $P_{\rm s}^{\rm MC}$ values are obtained using permutations with $10^5$ MC iterations.}
\end{minipage}
\end{table}

In this context, it should be taken into account that a significant correlations
between SNe~Ia decline rates (stretch parameters) and the global ages of hosts
have been observed when the ages range from about 1 to $\sim10$~Gyr
(e.g. \citealt{2011ApJ...740...92G,2014MNRAS.438.1391P,2016MNRAS.457.3470C};
\citetalias{2020MNRAS.499.1424H}).
In the stacked discs of Sa--Scd galaxies, however, the azimuthally averaged age of the stellar population
ranges roughly from 8.5 to 10~Gyr from the disc edge to the center, respectively
\citep[e.g.][]{2015A&A...581A.103G}.
Most likely, this narrow average age distribution across the mean (stacked) host disc
does not allow to see a significant correlation between
the $\Delta m_{15}$ and $\widetilde{r}_{\rm SN}$ in Table~\ref{SpRcordm15vsR}.

It is clear that such a mean disc contains an overlaid components of old and young stars at any radius.
On the other hand, as shown in \citet{1994ApJ...423L..31D,2016MNRAS.459.3130A},
a considerable fraction of SNe~Ia in spiral galaxies is (observationally) linked
to the young/star-forming disc population, rather than to the population of old disc or bulge.
These SNe~Ia exhibit an average delay time of $200-500$~Myr
\citep[prompt events, e.g.][]{2009ApJ...707...74R}
and should have slower declining LCs
\citep[smaller $\Delta m_{15}$ values, e.g.][]{2017ApJ...851L..50S}.
For this reason, the SNe~Ia host disc is outnumbered by slower declining events
outside the SFD (Fig.~\ref{dm15rsnrdemcum}).

Given these results, we check the $\Delta m_{15}$ -- $\widetilde{r}_{\rm SN}/\widetilde{r}_{\rm dem}$
correlation for normal SNe~Ia in the SFD$+$outer disc, bar/SF$+$outer disc, and combined samples.
Table~\ref{SpRcordm15vsR} shows that the mentioned correlation is statistically significant
for the first sample, while it is not significant for the second and combined samples.
Thus, the old SFD population ($\gtrsim$ 2~Gyr), which contains mostly faster declining SNe~Ia
(larger $\Delta m_{15}$), in combination with the younger outer disc,
which is outnumbered by SNe~Ia with slower declining LCs
(smaller $\Delta m_{15}$), cause the observed trend in
the SFD$+$outer disc (Fig.~\ref{dm15rsnrdemcum} and Table~\ref{SpRcordm15vsR}).

\section{Conclusions}
\label{DISconc}

In this \emph{Letter}, using a sample of nearby
Sa--Scd galaxies hosting 185 SNe~Ia and our
visual classification of the ionized (UV and/or H$\alpha$) discs of the galaxies,
we perform an analysis of the locations and LC decline rates
$(\Delta m_{15})$ of normal and peculiar SNe~Ia in the SFDs and beyond.
As in earlier studies, we confirm that in the stacked spiral disc,
the $\Delta m_{15}$ of SNe~Ia do not correlate with their galactocentric radii,
and such disc is outnumbered by slower declining/prompt events.
For the first time, we demonstrate that from the perspective of
the dynamical timescale of the SFD, its old stellar population
($\gtrsim$2~Gyr) hosts mostly faster declining SNe~Ia $(\Delta m_{15} > 1.25)$.
By linking the LC decline rate and progenitor age,
we show that the SFD phenomenon gives an excellent possibility to
constrain the nature of SNe~Ia.
We encourage further analysis (e.g. integral field observations)
using the SFD phenomenon on larger datasets of
SNe~Ia and their host galaxies to better constrain SN~Ia progenitor ages.

\section*{Acknowledgements}

We thank the anonymous referee for thoughtful comments and
efforts towards improving our \emph{Letter}.
This work was made possible in part by a research grant from
the Armenian National Science and Education Fund (ANSEF) based in New York, USA.
V.A. is supported by FCT through national grants:
UID/FIS/04434/2019; UIDB/04434/2020; UIDP/04434/2020,
and Investigador contract nr.~IF/00650/2015/CP1273/CT0001.


\section*{Data Availability}

The data underlying this study are available in the \emph{Letter},
in its online supplementary material (for guidance see Table~\ref{databasetab}),
and in \citetalias{2020MNRAS.499.1424H}.


\bibliography{SNIaSFDbib}

\newcommand{\SortNoop}[1]{}
\begin{thebibliography}{}
\makeatletter
\relax
\def\mn@urlcharsother{\let\do\@makeother \do\$\do\&\do\#\do\^\do\_\do\%\do\~}
\def\mn@doi{\begingroup\mn@urlcharsother \@ifnextchar [ {\mn@doi@}
  {\mn@doi@[]}}
\def\mn@doi@[#1]#2{\def\@tempa{#1}\ifx\@tempa\@empty \href
  {http://dx.doi.org/#2} {doi:#2}\else \href {http://dx.doi.org/#2} {#1}\fi
  \endgroup}
\def\mn@eprint#1#2{\mn@eprint@#1:#2::\@nil}
\def\mn@eprint@arXiv#1{\href {http://arxiv.org/abs/#1} {{\tt arXiv:#1}}}
\def\mn@eprint@dblp#1{\href {http://dblp.uni-trier.de/rec/bibtex/#1.xml}
  {dblp:#1}}
\def\mn@eprint@#1:#2:#3:#4\@nil{\def\@tempa {#1}\def\@tempb {#2}\def\@tempc
  {#3}\ifx \@tempc \@empty \let \@tempc \@tempb \let \@tempb \@tempa \fi \ifx
  \@tempb \@empty \def\@tempb {arXiv}\fi \@ifundefined
  {mn@eprint@\@tempb}{\@tempb:\@tempc}{\expandafter \expandafter \csname
  mn@eprint@\@tempb\endcsname \expandafter{\@tempc}}}

\bibitem[\protect\citeauthoryear{{Anderson}, {James}, {F{\"o}rster},
  {Gonz{\'a}lez-Gait{\'a}n}, {Habergham}, {Hamuy}  \& {Lyman}}{{Anderson}
  et~al.}{2015}]{2015MNRAS.448..732A}
{Anderson} J.~P.,  {James} P.~A.,  {F{\"o}rster} F.,  {Gonz{\'a}lez-Gait{\'a}n}
  S.,  {Habergham} S.~M.,  {Hamuy} M.,   {Lyman} J.~D.,  2015, \mn@doi [\mnras]
  {10.1093/mnras/stu2712}, \href
  {https://ui.adsabs.harvard.edu/abs/2015MNRAS.448..732A} {448, 732}

\bibitem[\protect\citeauthoryear{{Aramyan} et~al.,}{{Aramyan}
  et~al.}{2016}]{2016MNRAS.459.3130A}
{Aramyan} L.~S.,  et~al., 2016, \mn@doi [\mnras] {10.1093/mnras/stw873}, \href
  {https://ui.adsabs.harvard.edu/abs/2016MNRAS.459.3130A} {459, 3130}

\bibitem[\protect\citeauthoryear{{Barkhudaryan}, {Hakobyan}, {Karapetyan},
  {Mamon}, {Kunth}, {Adibekyan}  \& {Turatto}}{{Barkhudaryan}
  et~al.}{2019}]{2019MNRAS.490..718B}
{Barkhudaryan} L.~V.,  {Hakobyan} A.~A.,  {Karapetyan} A.~G.,  {Mamon} G.~A.,
  {Kunth} D.,  {Adibekyan} V.,   {Turatto} M.,  2019, \mn@doi [\mnras]
  {10.1093/mnras/stz2585}, \href
  {https://ui.adsabs.harvard.edu/abs/2019MNRAS.490..718B} {490, 718}

\bibitem[\protect\citeauthoryear{{Blondin}, {Dessart}, {Hillier}  \&
  {Khokhlov}}{{Blondin} et~al.}{2017}]{2017MNRAS.470..157B}
{Blondin} S.,  {Dessart} L.,  {Hillier} D.~J.,   {Khokhlov} A.~M.,  2017,
  \mn@doi [\mnras] {10.1093/mnras/stw2492}, \href
  {https://ui.adsabs.harvard.edu/abs/2017MNRAS.470..157B} {470, 157}

\bibitem[\protect\citeauthoryear{{Campbell}, {Fraser}  \& {Gilmore}}{{Campbell}
  et~al.}{2016}]{2016MNRAS.457.3470C}
{Campbell} H.,  {Fraser} M.,   {Gilmore} G.,  2016, \mn@doi [\mnras]
  {10.1093/mnras/stw115}, \href
  {https://ui.adsabs.harvard.edu/abs/2016MNRAS.457.3470C} {457, 3470}

\bibitem[\protect\citeauthoryear{{Childress}, {Wolf}  \& {Zahid}}{{Childress}
  et~al.}{2014}]{2014MNRAS.445.1898C}
{Childress} M.~J.,  {Wolf} C.,   {Zahid} H.~J.,  2014, \mn@doi [\mnras]
  {10.1093/mnras/stu1892}, \href
  {https://ui.adsabs.harvard.edu/abs/2014MNRAS.445.1898C} {445, 1898}

\bibitem[\protect\citeauthoryear{{Crocker} et~al.,}{{Crocker}
  et~al.}{2017}]{2017NatAs...1E.135C}
{Crocker} R.~M.,  et~al., 2017, \mn@doi [Nature Astronomy]
  {10.1038/s41550-017-0135}, \href
  {https://ui.adsabs.harvard.edu/abs/2017NatAs...1E.135C} {1, 0135}

\bibitem[\protect\citeauthoryear{{\SortNoop{Della Valle}}della~Valle \&
  {Livio}}{{\SortNoop{Della Valle}}della~Valle \&
  {Livio}}{1994}]{1994ApJ...423L..31D}
{\SortNoop{Della Valle}}della~Valle M.,  {Livio} M.,  1994, \mn@doi [\apjl]
  {10.1086/187228}, \href
  {https://ui.adsabs.harvard.edu/abs/1994ApJ...423L..31D} {423, L31}

\bibitem[\protect\citeauthoryear{{Di Matteo}, {Haywood}, {Combes}, {Semelin}
  \& {Snaith}}{{Di Matteo} et~al.}{2013}]{2013A&A...553A.102D}
{Di Matteo} P.,  {Haywood} M.,  {Combes} F.,  {Semelin} B.,   {Snaith} O.~N.,
  2013, \mn@doi [\aap] {10.1051/0004-6361/201220539}, \href
  {https://ui.adsabs.harvard.edu/abs/2013A&A...553A.102D} {553, A102}

\bibitem[\protect\citeauthoryear{{D{\'\i}az-Garc{\'\i}a}, {Salo}, {Laurikainen}
   \& {Herrera-Endoqui}}{{D{\'\i}az-Garc{\'\i}a}
  et~al.}{2016}]{2016A&A...587A.160D}
{D{\'\i}az-Garc{\'\i}a} S.,  {Salo} H.,  {Laurikainen} E.,   {Herrera-Endoqui}
  M.,  2016, \mn@doi [\aap] {10.1051/0004-6361/201526161}, \href
  {https://ui.adsabs.harvard.edu/abs/2016A&A...587A.160D} {587, A160}

\bibitem[\protect\citeauthoryear{{D{\'\i}az-Garc{\'\i}a}, {Moyano},
  {Comer{\'o}n}, {Knapen}, {Salo}  \& {Bouquin}}{{D{\'\i}az-Garc{\'\i}a}
  et~al.}{2020}]{2020A&A...644A..38D}
{D{\'\i}az-Garc{\'\i}a} S.,  {Moyano} F.~D.,  {Comer{\'o}n} S.,  {Knapen}
  J.~H.,  {Salo} H.,   {Bouquin} A.~Y.~K.,  2020, \mn@doi [\aap]
  {10.1051/0004-6361/202039162}, \href
  {https://ui.adsabs.harvard.edu/abs/2020A&A...644A..38D} {644, A38}

\bibitem[\protect\citeauthoryear{{Donohoe-Keyes}, {Martig}, {James}  \&
  {Kraljic}}{{Donohoe-Keyes} et~al.}{2019}]{2019MNRAS.489.4992D}
{Donohoe-Keyes} C.~E.,  {Martig} M.,  {James} P.~A.,   {Kraljic} K.,  2019,
  \mn@doi [\mnras] {10.1093/mnras/stz2474}, \href
  {https://ui.adsabs.harvard.edu/abs/2019MNRAS.489.4992D} {489, 4992}

\bibitem[\protect\citeauthoryear{{Fisher} \& {Jumper}}{{Fisher} \&
  {Jumper}}{2015}]{2015ApJ...805..150F}
{Fisher} R.,  {Jumper} K.,  2015, \mn@doi [\apj] {10.1088/0004-637X/805/2/150},
  \href {https://ui.adsabs.harvard.edu/abs/2015ApJ...805..150F} {805, 150}

\bibitem[\protect\citeauthoryear{{Galbany} et~al.,}{{Galbany}
  et~al.}{2012}]{2012ApJ...755..125G}
{Galbany} L.,  et~al., 2012, \mn@doi [\apj] {10.1088/0004-637X/755/2/125},
  \href {https://ui.adsabs.harvard.edu/abs/2012ApJ...755..125G} {755, 125}

\bibitem[\protect\citeauthoryear{{Gallagher}, {Garnavich}, {Berlind},
  {Challis}, {Jha}  \& {Kirshner}}{{Gallagher}
  et~al.}{2005}]{2005ApJ...634..210G}
{Gallagher} J.~S.,  {Garnavich} P.~M.,  {Berlind} P.,  {Challis} P.,  {Jha} S.,
    {Kirshner} R.~P.,  2005, \mn@doi [\apj] {10.1086/491664}, \href
  {https://ui.adsabs.harvard.edu/abs/2005ApJ...634..210G} {634, 210}

\bibitem[\protect\citeauthoryear{{George}, {Joseph}, {Mondal}, {Subramanian},
  {Subramaniam}  \& {Paul}}{{George} et~al.}{2020}]{2020A&A...644A..79G}
{George} K.,  {Joseph} P.,  {Mondal} C.,  {Subramanian} S.,  {Subramaniam} A.,
   {Paul} K.~T.,  2020, \mn@doi [\aap] {10.1051/0004-6361/202038810}, \href
  {https://ui.adsabs.harvard.edu/abs/2020A&A...644A..79G} {644, A79}

\bibitem[\protect\citeauthoryear{{Gonz{\'a}lez Delgado} et~al.,}{{Gonz{\'a}lez
  Delgado} et~al.}{2015}]{2015A&A...581A.103G}
{Gonz{\'a}lez Delgado} R.~M.,  et~al., 2015, \mn@doi [\aap]
  {10.1051/0004-6361/201525938}, \href
  {https://ui.adsabs.harvard.edu/abs/2015A&A...581A.103G} {581, A103}

\bibitem[\protect\citeauthoryear{{Gupta} et~al.,}{{Gupta}
  et~al.}{2011}]{2011ApJ...740...92G}
{Gupta} R.~R.,  et~al., 2011, \mn@doi [\apj] {10.1088/0004-637X/740/2/92},
  \href {https://ui.adsabs.harvard.edu/abs/2011ApJ...740...92G} {740, 92}

\bibitem[\protect\citeauthoryear{{Hakobyan} et~al.,}{{Hakobyan}
  et~al.}{2016}]{2016MNRAS.456.2848H}
{Hakobyan} A.~A.,  et~al., 2016, \mn@doi [\mnras] {10.1093/mnras/stv2853},
  \href {https://ui.adsabs.harvard.edu/abs/2016MNRAS.456.2848H} {456, 2848}

\bibitem[\protect\citeauthoryear{{Hakobyan}, {Barkhudaryan}, {Karapetyan},
  {Gevorgyan}, {Mamon}, {Kunth}, {Adibekyan}  \& {Turatto}}{{Hakobyan}
  et~al.}{2020}]{2020MNRAS.499.1424H}
{Hakobyan} A.~A.,  {Barkhudaryan} L.~V.,  {Karapetyan} A.~G.,  {Gevorgyan}
  M.~H.,  {Mamon} G.~A.,  {Kunth} D.,  {Adibekyan} V.,   {Turatto} M.,  2020,
  \mn@doi [\mnras] {10.1093/mnras/staa2940}, \href
  {https://ui.adsabs.harvard.edu/abs/2020MNRAS.499.1424H} {499, 1424 (H20)}

\bibitem[\protect\citeauthoryear{{Han} \& {Podsiadlowski}}{{Han} \&
  {Podsiadlowski}}{2004}]{2004MNRAS.350.1301H}
{Han} Z.,  {Podsiadlowski} P.,  2004, \mn@doi [\mnras]
  {10.1111/j.1365-2966.2004.07713.x}, \href
  {https://ui.adsabs.harvard.edu/abs/2004MNRAS.350.1301H} {350, 1301}

\bibitem[\protect\citeauthoryear{{James} \& {Percival}}{{James} \&
  {Percival}}{2015}]{2015MNRAS.450.3503J}
{James} P.~A.,  {Percival} S.~M.,  2015, \mn@doi [\mnras]
  {10.1093/mnras/stv846}, \href
  {https://ui.adsabs.harvard.edu/abs/2015MNRAS.450.3503J} {450, 3503}

\bibitem[\protect\citeauthoryear{{James} \& {Percival}}{{James} \&
  {Percival}}{2018}]{2018MNRAS.474.3101J}
{James} P.~A.,  {Percival} S.~M.,  2018, \mn@doi [\mnras]
  {10.1093/mnras/stx2990}, \href
  {https://ui.adsabs.harvard.edu/abs/2018MNRAS.474.3101J} {474, 3101}

\bibitem[\protect\citeauthoryear{{Kang}, {Lee}, {Kim}, {Chung}  \&
  {Ree}}{{Kang} et~al.}{2020}]{2020ApJ...889....8K}
{Kang} Y.,  {Lee} Y.-W.,  {Kim} Y.-L.,  {Chung} C.,   {Ree} C.~H.,  2020,
  \mn@doi [\apj] {10.3847/1538-4357/ab5afc}, \href
  {https://ui.adsabs.harvard.edu/abs/2020ApJ...889....8K} {889, 8}

\bibitem[\protect\citeauthoryear{{Kasen}, {R{\"o}pke}  \& {Woosley}}{{Kasen}
  et~al.}{2009}]{2009Natur.460..869K}
{Kasen} D.,  {R{\"o}pke} F.~K.,   {Woosley} S.~E.,  2009, \mn@doi [\nat]
  {10.1038/nature08256}, \href
  {https://ui.adsabs.harvard.edu/abs/2009Natur.460..869K} {460, 869}

\bibitem[\protect\citeauthoryear{{Kennicutt}}{{Kennicutt}}{1998}]{1998ARA&A..36..189K}
{Kennicutt} Robert~C. J.,  1998, \mn@doi [\araa]
  {10.1146/annurev.astro.36.1.189}, \href
  {https://ui.adsabs.harvard.edu/abs/1998ARA&A..36..189K} {36, 189}

\bibitem[\protect\citeauthoryear{{Livio} \& {Mazzali}}{{Livio} \&
  {Mazzali}}{2018}]{2018PhR...736....1L}
{Livio} M.,  {Mazzali} P.,  2018, \mn@doi [\physrep]
  {10.1016/j.physrep.2018.02.002}, \href
  {https://ui.adsabs.harvard.edu/abs/2018PhR...736....1L} {736, 1}

\bibitem[\protect\citeauthoryear{{Martin} et~al.,}{{Martin}
  et~al.}{2005}]{2005ApJ...619L...1M}
{Martin} D.~C.,  et~al., 2005, \mn@doi [\apjl] {10.1086/426387}, \href
  {https://ui.adsabs.harvard.edu/abs/2005ApJ...619L...1M} {619, L1}

\bibitem[\protect\citeauthoryear{{Minchev} et~al.,}{{Minchev}
  et~al.}{2018}]{2018MNRAS.481.1645M}
{Minchev} I.,  et~al., 2018, \mn@doi [\mnras] {10.1093/mnras/sty2033}, \href
  {https://ui.adsabs.harvard.edu/abs/2018MNRAS.481.1645M} {481, 1645}

\bibitem[\protect\citeauthoryear{{Pan} et~al.,}{{Pan}
  et~al.}{2014}]{2014MNRAS.438.1391P}
{Pan} Y.~C.,  et~al., 2014, \mn@doi [\mnras] {10.1093/mnras/stt2287}, \href
  {https://ui.adsabs.harvard.edu/abs/2014MNRAS.438.1391P} {438, 1391}

\bibitem[\protect\citeauthoryear{{Panther}, {Seitenzahl}, {Ruiter}, {Crocker},
  {Lidman}, {Wang}, {Tucker}  \& {Groves}}{{Panther}
  et~al.}{2019}]{2019PASA...36...31P}
{Panther} F.~H.,  {Seitenzahl} I.~R.,  {Ruiter} A.~J.,  {Crocker} R.~M.,
  {Lidman} C.,  {Wang} E.~X.,  {Tucker} B.~E.,   {Groves} B.,  2019, \mn@doi
  [\pasa] {10.1017/pasa.2019.24}, \href
  {https://ui.adsabs.harvard.edu/abs/2019PASA...36...31P} {36, e031}

\bibitem[\protect\citeauthoryear{{Pavlyuk} \& {Tsvetkov}}{{Pavlyuk} \&
  {Tsvetkov}}{2016}]{2016AstL...42..495P}
{Pavlyuk} N.~N.,  {Tsvetkov} D.~Y.,  2016, \mn@doi [Astron. Lett.]
  {10.1134/S1063773716080053}, \href
  {https://ui.adsabs.harvard.edu/abs/2016AstL...42..495P} {42, 495}

\bibitem[\protect\citeauthoryear{{Phillips}}{{Phillips}}{1993}]{1993ApJ...413L.105P}
{Phillips} M.~M.,  1993, \mn@doi [\apjl] {10.1086/186970}, \href
  {https://ui.adsabs.harvard.edu/abs/1993ApJ...413L.105P} {413, L105}

\bibitem[\protect\citeauthoryear{{Raskin}, {Scannapieco}, {Rhoads}  \& {Della
  Valle}}{{Raskin} et~al.}{2009}]{2009ApJ...707...74R}
{Raskin} C.,  {Scannapieco} E.,  {Rhoads} J.,   {Della Valle} M.,  2009,
  \mn@doi [\apj] {10.1088/0004-637X/707/1/74}, \href
  {https://ui.adsabs.harvard.edu/abs/2009ApJ...707...74R} {707, 74}

\bibitem[\protect\citeauthoryear{{Rigault} et~al.,}{{Rigault}
  et~al.}{2013}]{2013A&A...560A..66R}
{Rigault} M.,  et~al., 2013, \mn@doi [\aap] {10.1051/0004-6361/201322104},
  \href {https://ui.adsabs.harvard.edu/abs/2013A&A...560A..66R} {560, A66}

\bibitem[\protect\citeauthoryear{{Roming} et~al.,}{{Roming}
  et~al.}{2005}]{2005SSRv..120...95R}
{Roming} P. W.~A.,  et~al., 2005, \mn@doi [\ssr] {10.1007/s11214-005-5095-4},
  \href {https://ui.adsabs.harvard.edu/abs/2005SSRv..120...95R} {120, 95}

\bibitem[\protect\citeauthoryear{{Rose}, {Garnavich}  \& {Berg}}{{Rose}
  et~al.}{2019}]{2019ApJ...874...32R}
{Rose} B.~M.,  {Garnavich} P.~M.,   {Berg} M.~A.,  2019, \mn@doi [\apj]
  {10.3847/1538-4357/ab0704}, \href
  {https://ui.adsabs.harvard.edu/abs/2019ApJ...874...32R} {874, 32}

\bibitem[\protect\citeauthoryear{{Ruiter} et~al.,}{{Ruiter}
  et~al.}{2013}]{2013MNRAS.429.1425R}
{Ruiter} A.~J.,  et~al., 2013, \mn@doi [\mnras] {10.1093/mnras/sts423}, \href
  {https://ui.adsabs.harvard.edu/abs/2013MNRAS.429.1425R} {429, 1425}

\bibitem[\protect\citeauthoryear{{S{\'a}nchez-Menguiano}
  et~al.,}{{S{\'a}nchez-Menguiano} et~al.}{2018}]{2018A&A...609A.119S}
{S{\'a}nchez-Menguiano} L.,  et~al., 2018, \mn@doi [\aap]
  {10.1051/0004-6361/201731486}, \href
  {https://ui.adsabs.harvard.edu/abs/2018A&A...609A.119S} {609, A119}

\bibitem[\protect\citeauthoryear{{Shen} \& {Sellwood}}{{Shen} \&
  {Sellwood}}{2004}]{2004ApJ...604..614S}
{Shen} J.,  {Sellwood} J.~A.,  2004, \mn@doi [\apj] {10.1086/382124}, \href
  {https://ui.adsabs.harvard.edu/abs/2004ApJ...604..614S} {604, 614}

\bibitem[\protect\citeauthoryear{{Shen}, {Toonen}  \& {Graur}}{{Shen}
  et~al.}{2017}]{2017ApJ...851L..50S}
{Shen} K.~J.,  {Toonen} S.,   {Graur} O.,  2017, \mn@doi [\apjl]
  {10.3847/2041-8213/aaa015}, \href
  {https://ui.adsabs.harvard.edu/abs/2017ApJ...851L..50S} {851, L50}

\bibitem[\protect\citeauthoryear{{Shen}, {Blondin}, {Kasen}, {Dessart},
  {Townsley}, {Boos}  \& {Hillier}}{{Shen} et~al.}{2021}]{2021ApJ...909L..18S}
{Shen} K.~J.,  {Blondin} S.,  {Kasen} D.,  {Dessart} L.,  {Townsley} D.~M.,
  {Boos} S.,   {Hillier} D.~J.,  2021, \mn@doi [\apjl]
  {10.3847/2041-8213/abe69b}, \href
  {https://ui.adsabs.harvard.edu/abs/2021ApJ...909L..18S} {909, L18}

\bibitem[\protect\citeauthoryear{{Sim}, {R{\"o}pke}, {Hillebrandt}, {Kromer},
  {Pakmor}, {Fink}, {Ruiter}  \& {Seitenzahl}}{{Sim}
  et~al.}{2010}]{2010ApJ...714L..52S}
{Sim} S.~A.,  {R{\"o}pke} F.~K.,  {Hillebrandt} W.,  {Kromer} M.,  {Pakmor} R.,
   {Fink} M.,  {Ruiter} A.~J.,   {Seitenzahl} I.~R.,  2010, \mn@doi [\apjl]
  {10.1088/2041-8205/714/1/L52}, \href
  {https://ui.adsabs.harvard.edu/abs/2010ApJ...714L..52S} {714, L52}

\bibitem[\protect\citeauthoryear{{Timmes}, {Brown}  \& {Truran}}{{Timmes}
  et~al.}{2003}]{2003ApJ...590L..83T}
{Timmes} F.~X.,  {Brown} E.~F.,   {Truran} J.~W.,  2003, \mn@doi [\apjl]
  {10.1086/376721}, \href
  {https://ui.adsabs.harvard.edu/abs/2003ApJ...590L..83T} {590, L83}

\bibitem[\protect\citeauthoryear{{Uddin}, {Mould}, {Lidman}, {Ruhlmann-Kleider}
   \& {Zhang}}{{Uddin} et~al.}{2017}]{2017ApJ...848...56U}
{Uddin} S.~A.,  {Mould} J.,  {Lidman} C.,  {Ruhlmann-Kleider} V.,   {Zhang}
  B.~R.,  2017, \mn@doi [\apj] {10.3847/1538-4357/aa8df7}, \href
  {https://ui.adsabs.harvard.edu/abs/2017ApJ...848...56U} {848, 56}

\bibitem[\protect\citeauthoryear{{Wang}, {H{\"o}flich}  \& {Wheeler}}{{Wang}
  et~al.}{1997}]{1997ApJ...483L..29W}
{Wang} L.,  {H{\"o}flich} P.,   {Wheeler} J.~C.,  1997, \mn@doi [\apjl]
  {10.1086/310737}, \href
  {https://ui.adsabs.harvard.edu/abs/1997ApJ...483L..29W} {483, L29}

\makeatother
\end{thebibliography}

\section*{Supporting information}

Supplementary data are available at \emph{MNRAS} online.\\
\\
\textbf{Table~\ref{databasetab}.}
The database of 185 individual SNe~Ia and their 180 host galaxies.\\
\\
Please note: Oxford University Press is not responsible for the
content or functionality of any supporting materials supplied by
the authors. Any queries (other than missing material) should be
directed to the corresponding author for the article.


\appendix
\section{Online material}

The database of our analysis is available in the online
supplementary material of the \emph{Letter}.
The first 10 rows of the database of 185 SNe~Ia
(SN name, location, deprojected and $R_{25}$-normalized galactocentric distance)
and their 180 hosts (galaxy name, morphological type, bar detection,
disc's class, and demarcation radius)
are shown in Table~\ref{databasetab}.
The full table is available in an CSV format.
Recall that more data on these SNe~Ia and their host galaxies are available
in \citetalias{2020MNRAS.499.1424H} (e.g. SN spectroscopic subclass,
$\Delta m_{15}$, galaxy distance).

\begin{table}
  \centering
  \begin{minipage}{110mm}
  \caption{The database (first 10 rows) of 185 SNe~Ia and their 180 host galaxies.
           The full table is available as supplementary material.}
  \tabcolsep 6pt
  \label{databasetab}
  \begin{tabular}{llllllcc}
    \hline
    \multicolumn{1}{c}{SN} & \multicolumn{1}{c}{Host} & \multicolumn{1}{c}{Morph.} & \multicolumn{1}{c}{Bar} & \multicolumn{1}{c}{Disc's class} & \multicolumn{1}{c}{Location} & \multicolumn{1}{c}{$\widetilde{r}_{\rm SN}$} & \multicolumn{1}{c}{$\widetilde{r}_{\rm dem}$}\\
    \hline
     1974G & NGC4414 & Sc &  & \emph{i} & disc & 0.425 & --\\
     1981B & NGC4536 & Sbc &  & \emph{i} & disc & 0.697 & --\\
     1982B & NGC2268 & Sc & B & \emph{iii}? & outer~disc & 0.267 & 0.150\\
     1989A & NGC3687 & Sc & B & \emph{iii}: & outer~disc & 0.524 & 0.175\\
     1989B & NGC3627 & Sb & B & \emph{ii} & bar/SF & 0.171 & 0.200\\
     1990N & NGC4639 & Sbc & B & \emph{iii} & outer~disc & 0.859 & 0.253\\
     1990O & MCG+03-44-003 & Sbc & B & \emph{iii}? & outer~disc & 0.764 & 0.333\\
     1991T & NGC4527 & Sbc &  & \emph{i} & disc & 0.518 & --\\
     1992bc & ESO300-009 & Scd: &  & \emph{i} & disc & 0.897 & --\\
     1992bg & PGC343503 & Sb: &  & \emph{i} & disc & 0.546 & --\\
    \hline
  \end{tabular}
  \end{minipage}
\end{table}

\begin{flushleft}
{\small This paper has been typeset from a {\TeX/\LaTeX} file prepared by the author.}
\end{flushleft}

\label{lastpage}

\end{document}